\documentclass[5p,10pt]{elsarticle}

\biboptions{sort&compress}

\usepackage{amsmath}
\usepackage{geometry}
\usepackage{graphicx}
\usepackage[colorlinks=true,linkcolor=magenta,citecolor=cyan]{hyperref}
\usepackage{booktabs}
\usepackage{tabularx}
\usepackage{placeins}

\journal{Current Opinion in Structural Biology}

\begin{document}
\begin{frontmatter}

\title{Machine learning methods to study sequence--ensemble--function relationships in disordered proteins}

\author[1]{S{\"o}ren von B{\"u}low\fnref{fn1}}
\author[1]{Giulio Tesei\fnref{fn1}}
\author[1]{Kresten Lindorff-Larsen\corref{cor1}}
\ead{lindorff@bio.ku.dk}

\cortext[cor1]{Corresponding author}
\fntext[fn1]{These authors contributed equally to this work.}

\affiliation[1]{organization={Structural Biology and NMR Laboratory \& the Linderstr{\o}m-Lang Centre for Protein Science, Department of Biology, University of Copenhagen},
addressline={Ole Maaløes Vej 5},
postcode={2200},
postcodesep={},
city={Copenhagen},
country={Denmark}}

\begin{abstract}
Recent years have seen tremendous developments in the use of machine learning models to link amino acid sequence, structure and function of folded proteins. These methods are, however, rarely applicable to the wide range of proteins and sequences that comprise intrinsically disordered regions. We here review developments in the study of sequence--ensemble--function relationships of disordered proteins that exploit or are used to train machine learning models. These include methods for generating conformational ensembles and designing new sequences, and for linking sequences to biophysical properties and biological functions. We highlight how these developments are built on a tight integration between experiment, theory and simulations, and account for evolutionary constraints, which operate on sequences of disordered regions differently than on those of folded domains.
\end{abstract}

\end{frontmatter}

\section*{Introduction}

Intrinsically disordered proteins and regions (collectively, IDRs) are found across all domains of life, where they are involved in a wide range of cellular processes such as signaling, transcription, and sub-cellular organization \citep{Wright2015, Holehouse2024}. 
Previous reviews have discussed IDRs, their identification \citep{Zhao2022, Erdos2024}, their conformational characterization \citep{Holehouse2024}, their role in physiology and disease \citep{Uversky2008, Babu2016}, their interacting domains (short linear motifs, SLiMs) \citep{Davey2023}, and their phase behavior \citep{Shin2017, Mittag2022, Saar2023}. While referring the interested reader to these detailed reviews, we here focus on the use of machine learning (ML) methods to study IDR sequence--ensemble--function relationships.

\begin{figure}[tbp]
\centering
\includegraphics[width=\linewidth]{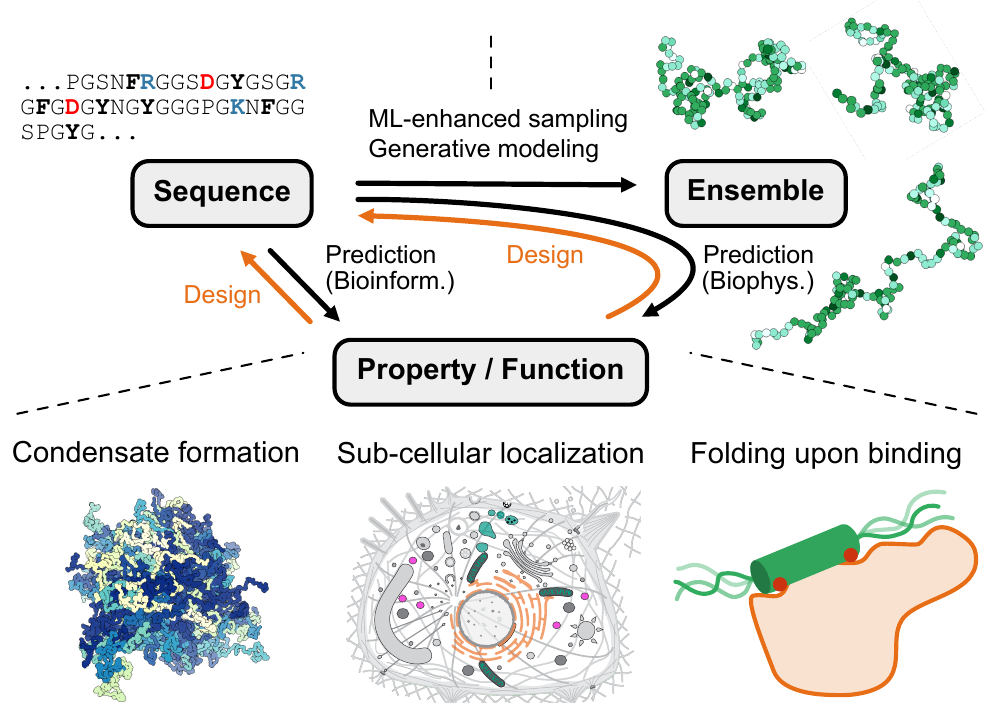}
\caption{Illustration of sequence--ensemble--function relationships for disordered proteins. ML prediction (black) and design (orange) approaches are highlighted on the connecting arrows. Prediction of properties/functions from sequence (or vice versa, design) can include biophysics approaches via structural ensembles, or bioinformatics approaches via other heterogeneous sources. The lower panels show examples of properties and functions of IDRs for predictions or design targets. The schematic illustration of an animal cell is adapted from the SwissBioPics template \citep{LeMercier2022}, under a Creative Commons licence CC BY 4.0.}
\label{fig1}
\end{figure}

ML is in the process of revolutionizing how we make sense of experimental and simulation data in science, including in the study of IDRs, and how we generate new data altogether \citep{Lindorff-Larsen2021, Ramanathan2021, Aupic2024}. ML methods, in particular neural networks (NN), are now routinely used to help understand the connections between IDR sequences, ensembles, biophysical properties, and biological functions (Figure~\ref{fig1}). 
To illustrate these developments, the following sections will highlight and review examples of recent advances in the use of ML to study IDR ensembles and function, recognizing that due to space constraints we cannot cover all the exciting work in this area. Notably, we do not discuss here the important work on identifying disordered protein regions using ML \citep{Tang2021, Hu2021b, Emenecker2021, Pang2023, Pang2024, Wang2024a, Nambiar2024}, as assessed by the critical assessment of protein intrinsic disorder (CAID) prediction challenge \citep{Conte2023}. Technical biophysics and ML terms are briefly explained in the Glossary (Table~\ref{tab:1}).

IDRs defy conventional sequence--structure--function relationships and instead usually exist as an ensemble of inter-converting configurations of similar free energy \citep{Lindorff-Larsen2021, Holehouse2024, Banerjee2024}. The structural diversity of the conformational ensembles and complex evolutionary signatures make IDRs particularly elusive for experimental and computational characterization. Nevertheless, experimental and simulation techniques 
have been used to determine the local and global structures and dynamics of IDRs as well as their interactions with binding partners \citep{Thomasen2022a, Orand2025}.
In concert, experiments, theory and simulations have been used to study conformational heterogeneity, phase behavior, and biological functions \citep{Lindorff-Larsen2021, Holehouse2024} of IDRs, but the set of rules connecting sequence features to structural ensembles are incompletely understood. ML methods have become important assets in understanding sequence--ensemble relationships both by increasing the accuracy of simulation models and by direct sampling via generative models, which we discuss in the first section of this review.

Likewise, the prediction of properties and functions of IDRs from sequence has been revolutionized by ML methods, as discussed in the second section. ML aids in the interpretation of structural ensemble data, and even ML models with modest complexity can successfully predict biophysical properties when trained on carefully parameterized physics-based models. We subsequently discuss how ML methods are in the process of unlocking our understanding of how IDRs shape protein function. Being primed for disentangling correlations and structure in data, ML methods help predicting IDR properties and functions like subcellular localization, phase separation, or binding to folded proteins. Furthermore, ML methods can help uncover evolutionary constraints on IDR sequences, sometimes circumventing limitations of classic alignment-based methods.

Finally, ML can aid in the design of IDRs with desired structural ensembles and/or properties, again via biophysics or bioinformatics approaches. Designing IDRs with specific properties---such as chain expansion, propensity to phase separate or to conditionally fold upon binding---is challenging because of the lack of one-to-one structure--function relationships. In the third section of this review, we discuss how ML methods are beginning to be used to overcome this challenge.

\section*{Computational sampling of IDR ensembles}

The precise knowledge of the structures constituting IDR ensembles makes it possible to characterize both the global dimensions and local transient structure of IDRs, which govern binding modes to interaction partners, spacing of folded domains, phase separation and aggregation \citep{Gonzalez-Foutel2022, Banerjee2024, Holehouse2024, Orand2025}. Further, accurate representation of IDR ensembles can predict effects of amino-acid substitutions and binding to small molecules \citep{Tesei2021, Banerjee2024, Dhar2025}.

Sampling the diverse ensemble of IDR configurations is challenging even for \textit{in vitro/in silico} systems of reduced complexity and heterogeneity compared to the cellular environment. ML has become a cornerstone in the efforts to computationally sample IDR ensembles, and has been applied via two approaches: ML-based parameterization of force fields for molecular simulations and generative modeling (Figure~\ref{fig1}).

Molecular dynamics (MD) simulations have been instrumental to generate conformational ensembles, study the dynamics of biomolecules, and link both structure and dynamics to biological functions. 
MD simulations numerically integrate equations of motion based on force fields that quantify interactions between atomistic or coarse-grained (CG) particles, depending on the desired system size, time scale, and computational costs. 
Atomistic MD simulations of IDRs in explicit solvent ideally generate accurate ensembles at high resolution with correct overall dimensions, transient secondary structure formation, intra- and intermolecular interactions, and other properties. However, despite various improvements to better represent both folded and disordered proteins \citep{Piana2015, Huang2017, Robustelli2018a, Piana2020}, atomistic MD simulations are still limited by force-field accuracy and the cost of performing simulations long enough for exhaustive sampling of the protein configurations \citep{Sarthak2023}.

Implicit solvent models or CG models help overcome the sampling problem by striking different balances between resolution and scalability \citep{Vitalis2009, Choi2019a, Dignon2018, Souza2021a}. 
Parameterizing these models requires optimization of force-field parameters against observables from atomistic MD simulations (bottom-up approach) like particle pair-distance distributions or experimental measurements like radii of gyration or saturation concentrations (top-down approach) \citep{Norgaard2008, DiPierro2013, Wang2013, Demerdash2019, Martin2020, Regy2021, Dannenhoffer-Lafage2021, Latham2021, Tesei2021, Farag2022, Valdes-Garcia2023, Ruff2015, Ding2024a, Souza2021a, Joseph2021, Jussupow2025}. 
The optimization problem is usually analytically intractable, but can be expressed as a supervised learning problem for ML.
One common strategy is to nudge model parameters, perform MD simulations, and re-assess the goodness of the model, e.g. the agreement to the experimental observables. To avoid over-fitting to the training data, the parameter optimization can be regularized \citep{Norgaard2008, Wang2013a, Ruff2015, Tesei2021, Ding2024a}.

\begin{figure*}[ht!]
\centering
\includegraphics[width=\linewidth]{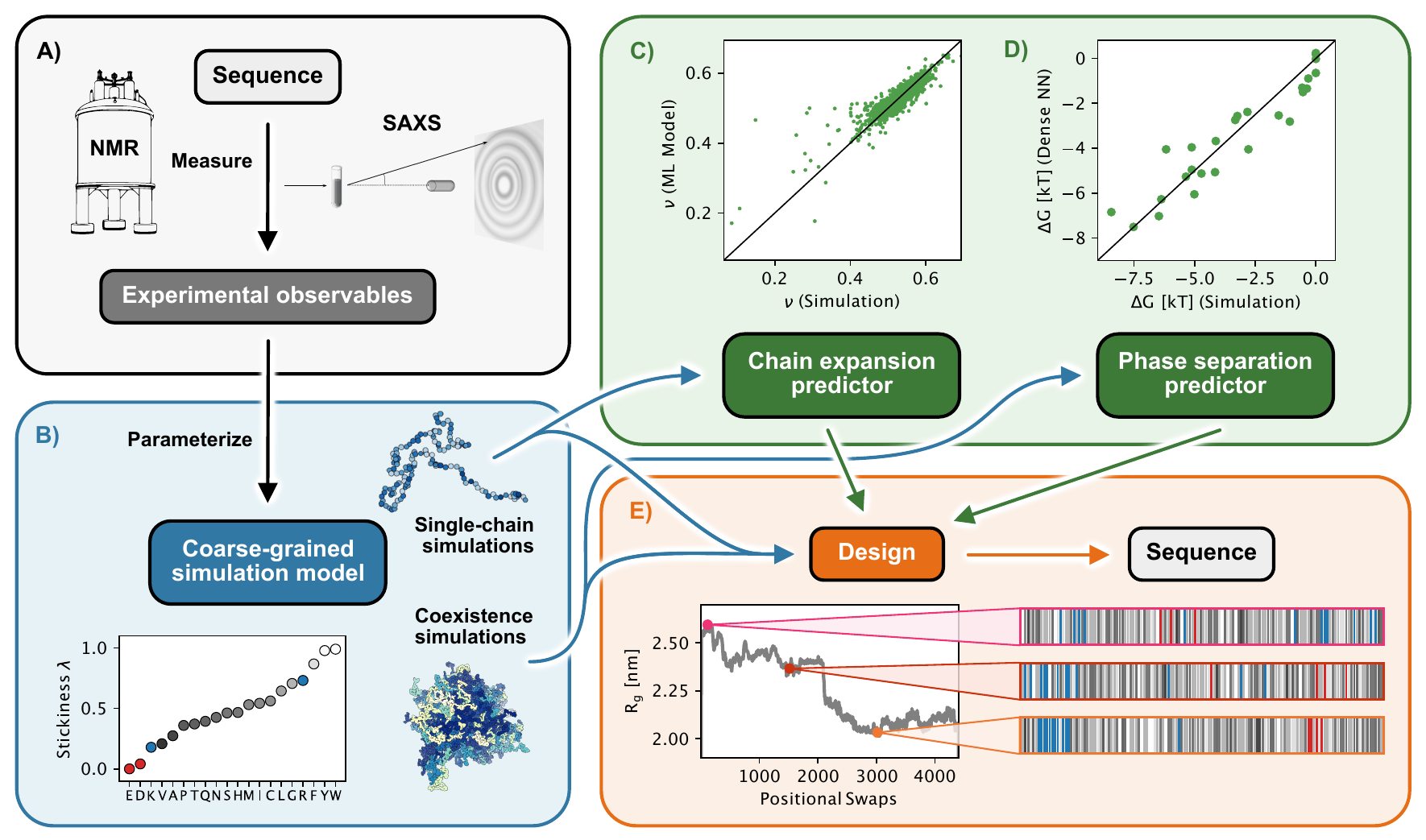}
\caption{Learning sequence--ensemble--property relationships from sparse data. (A--B) Recent studies have demonstrated how information from limited experimental data can be combined with physical models using ML to build accurate and fast coarse-grained simulation models \citep{Tesei2021, Tesei2023}. The graph in (B) shows the `stickiness' parameters $\lambda$ of the CALVADOS 2 model \citep{Tesei2023}. (C--D) Single-chain and direct-coexistence simulations can provide training data to inform ML predictors on complex relationships between sequence, chain compaction ($\nu$, the apparent scaling exponent), and phase separation propensities of IDRs \citep{Tesei2024,Lotthammer2024,vonBulow2024}, beyond the sequence scope of the original experimental data. (E) The simulation model and downstream ML models can be used to guide the design of IDRs with target properties. The example graph shows the progressive decrease of the radius of gyration, $R_g$, during the design of sequences with targeted high chain compaction using swap moves at fixed amino acid composition \citep{Pesce2024}. Vertical stripes represent the amino acid sequence for three intermediate solutions of the design procedure, using the same color code as for the stickiness scale in (B).}
\label{fig2}
\end{figure*}

As one example, we introduced a top-down Bayesian parameter-learning procedure to optimize parameters for IDR CG models against experimental observables \citep{Norgaard2008, Tesei2021, Tesei2023}(Figure~\ref{fig2}A,B). 
Here, the strength of non-electrostatic interactions between protein particles is encoded in a residue type-specific stickiness parameter $\lambda$ that relates to hydrophobicity scales. 
We targeted experimental SAXS and NMR data to optimize $\lambda$, and used previously published hydrophobicity scales as a Bayesian prior to reduce overfitting  \citep{Tesei2021,Tesei2023}. The resulting data-driven CALVADOS models \citep{Tesei2021,Tesei2023,Cao2024} thereby capture global conformational properties of IDRs and multi-domain proteins under common solution conditions across a relatively large region of sequence space \citep{Tesei2024,Pesce2024}.

Generative ML models replace simulation-based structure sampling altogether and are trained on simulation and/or experimental data \citep{Janson2023, Janson2024, Zhang2023, Lewis2024}.
For folded proteins, modern ML algorithms such as AlphaFold leverage the distance information encoded in evolutionary signals (e.g. via multiple-sequence alignment) to predict protein structures \citep{Jumper2021,Baek2021,Abramson2024} and thermodynamic ensembles from sequence \citep{Rotskoff2024}, or to design novel protein folds \citep{Anishchenko2021, Lin2024}. Whereas a small number of structures may be representative for highly populated states in folded proteins and informative for function, highly dynamic IDR ensembles are characterized by a large set of diverse conformations \citep{Lindorff-Larsen2021,Ruff2021}. Moreover, the evolutionary restraints that do operate on the highly variable sequences of IDRs are not easily picked up by procedures like multiple-sequence alignment. Nevertheless, recent work has shown some success in using AlphaFold-Multimer for predicting IDR binding modes to folded proteins \citep{Omidi2024}, as discussed in the next section.

Overall, the rapid prediction of accurate conformational ensembles of IDRs is a difficult problem \citep{Ruff2021}. The last years have seen a flurry of activity to generate models of IDR ensembles from sequence. Fragment-based methods stitch together exhaustively sampled peptide fragments \citep{Ozenne2012, Pietrek2020, Teixeira2022, Liu2023}. In contrast, ML-based methods learn the characteristics of the full IDR chain ensembles from molecular simulations and experimental data.

In one set of ML approaches, protein-specific models generate additional structures based on a set of pre-existing structures of the same or a closely related sequence, e.g. from short MD simulations \citep{Gupta2022, Taneja2024, Zhu2024}. These models can be seen as a form of ML-enhanced sampling of conformations. Another protein-specific approach is illustrated with the generative recurrent NN DynamICE, which aims to generate structural ensembles that are consistent with experimental NMR data \citep{Zhang2023}. Rather than reweighting conformational ensembles to better match the experimental data, DynamICE modulates the structures of existing ensembles via iterative updates of the residue torsion angles \citep{Zhang2023}. It thus biases the configurations using experimental constraints
even for scenarios with limited overlap between the original ensemble and experimental data.

An ideal generative IDR ensemble model would quickly and accurately generate structural ensembles for any IDR solely from sequence and given conditions such as temperature and pH. 
Recently, progress has been made in this endeavour, in part thanks to advances in the application of novel generative NN architectures.
The NNs are trained on a set of IDR simulation trajectories and are designed to be transferable to generate structures of previously unseen IDR sequences. IdpGAN uses a generative adversarial network (GAN) for fast and relatively accurate sampling of CG ensembles, albeit with lower transferability for atomistic modelling \citep{Janson2023}.
A new generation of networks such as idpSAM \citep{Janson2024} and IDPFold \citep{Zhu2024b} is based on diffusion models. Diffusion models sequentially add noise to input data in a multi-step forward process and learn the applied transformations in the reverse process \citep{Ho2020}. These models drastically improved the transferability for atomistic ensembles at the cost of reduced structure generation speed. 

Naturally, the accuracy of the ML-generated conformational ensemble depends on the input training data. Currently, these generative models are mostly trained on simulation data and thereby inherit the limitations and structural biases of the various molecular models used to generate the training data. Furthermore, the dynamics of MD simulations are generally not captured by generative models that draw ensembles in a probabilistic fashion.

Interpretation of the ensembles sampled with molecular force fields (aided by ML) or generative models and comparison with experimental data is crucial to ensure that the properties calculated from the ensembles are predictive and generalizable.
Quantitative comparison with experimental data often requires atomistic resolution, e.g. for calculating SAXS curves and NMR data using well-established routines \citep{Thomasen2022}. Likewise, ensemble docking approaches to small molecules require atomistic resolution \citep{Dhar2025}. When working with CG models, these and other tasks can be performed using backmapping procedures that estimate all-atom structures from lower-resolution representations \citep{Wassenaar2014}. Recently, a ML approach combined side-chain rotamer libraries with a transformer NN architecture to reconstruct side-chain rotamer states in the context of the full IDR \citep{Chennakesavalu2024}. In cg2all \citep{Heo2024}, a graph NN reconstructs atomistic detail from alpha-carbon representations and residue-type information and may considerably speed up generation of realistic all-atom structures from CG representations.

\section*{Predicting IDR properties and functions from sequence}

\begin{figure}[tbp]
 \centering
 \includegraphics[width=\columnwidth]{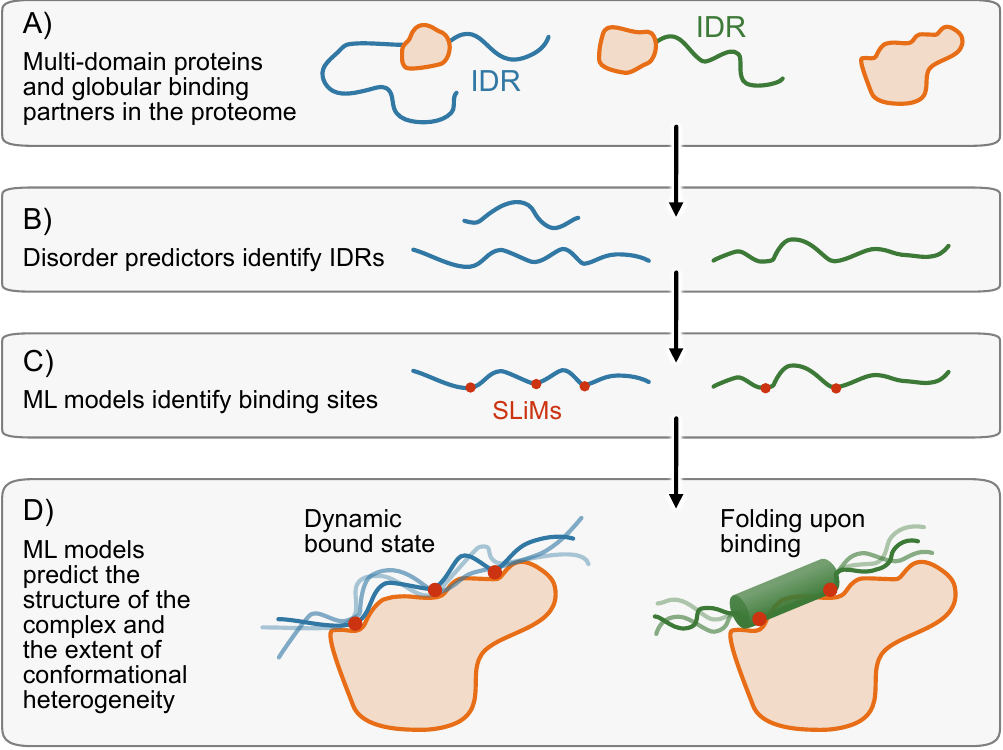} \caption{ML models in workflows to study interactions of IDRs with their binding partners. (A) Deep-learning models provide protein structures at the proteome scale \citep{Jumper2021,Baek2021,Abramson2024}. (B) Disorder predictors are used to identify IDRs from full-length sequences \citep{Conte2023}. (C) Approaches based on AlphaFold-Multimer and pLMs identify binding regions \citep{Jahn2024,Bret2024,Omidi2024}. (D) AlphaFold-Multimer predicts the binding modes of IDRs that fold upon binding and distinguishes between fuzzy and less dynamic protein complexes \citep{Omidi2024,Wu2024}.}
 \label{fig3}
 \end{figure}

The conformational properties of IDRs, which we have focused on above, are important determinants of protein function. How expanded or compact an IDR is can affect the degree to which SLiMs and other interaction-prone features are exposed to binding partners. For example, mutations that expand the global dimensions of transcription factors have been correlated with increased transcriptional activity \citep{Flores2025}. Moreover, the molecular interactions underpinning single-chain compaction largely overlap with the driving forces for homotypic phase separation, which suggests that biophysical properties of IDRs can inform us on the function of the full-length protein, including its involvement in the formation of biomolecular condensates \citep{Martin2020}. ML approaches can be used at multiple steps in linking IDR sequence to function including identifying disordered regions, locating linear motifs and predicting properties of functional complexes (Fig.~\ref{fig3}).
We here discuss approaches that leverage either biophysical or evolutionary information to study the relationship between amino acid sequences and the properties and functions of IDRs.

In biophysical ML approaches, training is performed on homogeneous datasets generated for example from CG simulations such as those described above. By capturing the physics governing for example conformational and phase separation properties of IDRs, CG molecular models can inform ML models, including fine-tuning of sequence-based ML models, that provide high-throughput sequence-based predictions \citep{Zheng2020,Lotthammer2024,Tesei2024,vonBulow2024,Seth2024,Mollaei2024} (Figure~\ref{fig2}C,D). Proteome-wide analyses have enabled systematic investigations of sequence--ensemble--function relationships, which highlighted associations between the overall dimensions of IDRs and the function and cellular localization of the full-length proteins \citep{Lotthammer2024,Tesei2024}. These large simulation datasets can then be used to train prediction models to calculate conformational properties directly from sequence, thus bypassing the need for simulations and considerably speeding up analyses, which can then scale to millions of sequences \citep{Lotthammer2024,Tesei2024} (Figure~\ref{fig2}C). 

These approaches can also be extended to studies of self-assembly and phase separation of IDRs. For example, we have trained a small dense NN on hundreds of simulations of two-phase systems with the CALVADOS model, using an active learning framework which guides the generation of training data in the most informative regions of sequence space. This ML model uses physically motivated features as input to provide rapid quantitative predictions of homotypic phase separation (single IDR components) that can be directly compared to experimental results and have accuracy comparable to computationally intensive simulations \citep{vonBulow2024} (Figure~\ref{fig2}D). The above studies demonstrate that even modest ML architectures can be used to accurately predict biophysical properties of IDRs from limited experimental data via two steps: (i) Train physics-based simulations models on sparse experimental data, and (ii) train ML prediction models on the larger simulation dataset.

Bioinformatics approaches have leveraged evolutionary conservation of IDR sequences or large, and often heterogeneous, experimental datasets. While it is clear that the biophysical, functional and sequence properties of many IDRs are conserved by evolution, it can be difficult to assess this evolutionary conservation directly through multiple-sequence alignments \citep{Lindorff-Larsen2021}. However, homologous IDRs can sometimes be identified as the regions flanked by conserved structured domains, which are instead more readily detected by alignment-based methods \citep{Toth-Petroczy2016} or possibly by assessment with alignment-free methods \citep{Chow2024}.

With a set of homologous IDRs, one can search for specific sequence properties that appear to be under selection during evolution and thus likely important for function. Statistical analyses of homologous IDRs have uncovered associations between specific biological functions and conserved sequence features, including linear motifs, net charge, and the patterning of charged and aromatic residues \citep{Zarin2017,Zarin2019}. These findings prompted the development of statistical models that link sequence features with protein-level functional annotations to infer which IDRs within a protein are likely responsible for specific functions \citep{Zarin2021,Singleton2024}. Further developments of the accuracy and scalability of these methods were exploited to generate proteome-wide predictions for a functional map that connects sequence features of all human IDRs with their function and cellular localization \citep{Pritisanac2024}.

The diversity of IDR sequence features highlights the importance of obtaining a comprehensive catalog based exclusively on evolutionary conservation, without biases from previous knowledge on functional annotations \citep{Lu2020,Lu2022b,Ho2023}. To achieve this goal, self-supervised contrastive learning was used to train NNs on large sets of homologous IDR sequences to learn features and specific residues that are conserved and functionally important \citep{Lu2020,Lu2022b,Ho2023}.
Additionally, ML models have shown success in identifying binding sites and characterizing the structure of IDR-protein complexes (Figure~\ref{fig3}). Protein language models (pLMs) trained on experimentally curated annotations enabled rapid predictions of binding regions within IDRs \citep{Jahn2024}. Approaches based on AlphaFold-Multimer also accurately identify binding sites in complexes of an IDR with a folded protein \citep{Bret2024}, and provide accurate structural predictions of IDR segments that fold upon binding \citep{Omidi2024}. Further, AlphaFold-Multimer confidence scores enable the classification of protein-IDR complexes as either conformationally homogeneous or fuzzy, where the structure of the IDR segment remains heterogeneous in the bound state. However, these models do not yet provide quantitative predictions of binding energetics or of the relative populations of conformers in the bound state \citep{Omidi2024}. Further, because the molecular context and nonspecific interactions of residues surrounding linear motifs impact IDR binding affinities \citep{Bugge2020,Langstein-Skora2022}, future developments may integrate bioinformatics approaches with physical models that account for 3D structure and interaction energies \citep{DelRosso2024}.

Curated annotations and proteomics data have also been instrumental to develop ML models that link amino acid sequences with homotypic phase separation \citep{Saar2021} and partitioning of IDRs into multi-component condensates \citep{Kilgore2025,Saar2024}. These models were tested by in-cell experiments, and are beginning to provide additional insights into the `molecular grammar' of phase separation in cells \citep{Kilgore2025,Saar2024}. Future work may combine such data with biophysical modelling and ML to provide further insight into the mechanisms that govern the existence of multiple independent condensates in cells. ML models such as those described above for calculating IDR properties such as single-chain compaction and phase separation can also be used to assess effects of missense variants \cite{Tesei2024,Kilgore2025}.

\section*{Designing IDRs with ML}

IDRs fall outside the scope of traditional structure-based protein design methods because their functions are not governed by single static structures. However, advances in molecular modeling of IDRs and deep-learning methods have recently opened new avenues for designing IDRs. One set of approaches leverages models that capture the physics underlying conformational and phase properties while providing computationally efficient predictions \citep{Harmon2016,Zeng2021,Lichtinger2021,Patel2024,vanHilten2023, An2024,Pesce2024,Changiarath2024,Krueger2024}. 

Simulation-based design approaches may be computationally costly, but can be made more efficient using ML methods \citep{Methorst2024}. As described above, large-scale simulations have been used to train models to predict IDR properties directly from sequence \citep{Zheng2020,Lotthammer2024,Tesei2024,vonBulow2024}, and such pre-trained models may be used as surrogates to drive design of IDRs with specified conformational or phase separation properties \citep{Pesce2024,Emenecker2023,vonBulow2024} (Figure~\ref{fig2}E). Alternatively, surrogate models may be refined iteratively during the design process through active learning frameworks \citep{Patel2024,An2024,Changiarath2024}. The active learning approach has been particularly effective in designing sequences that form biomolecular condensates, where one challenge is to balance thermodynamic stability with high internal dynamics \citep{An2024}. In this process, multi-objective Bayesian optimization was used to identify sequences that satisfy the trade-off between competing goals, such as the strength of self-interactions and the diffusion coefficient of an IDR in the dense phase \citep{Patel2024,An2024}. 

Bayesian optimization has also been used in the design of multiphase condensates and peptides that partition into preformed condensates \citep{Changiarath2024}. Here, the surrogate model was an ensemble of NNs, trained on data generated from CG simulations. 
By balancing exploration of untested sequences with exploitation of known successful candidates, the function prioritized sequences for further simulation, improving the accuracy of the NNs while approaching the optimization target \citep{Changiarath2024}.

In contrast to physics-based models, deep-learning generative sequence models offer a faster and potentially more versatile approach to IDR design, although they may lack the ability of physics-based models of tuning specific interactions. By focusing exclusively on sequence data, these models bypass the need for predefined structural templates, which are generally unavailable for IDRs. Sequence-based generative models, such as EvoDiff, use evolutionary-scale datasets to predict and generate sequences that fulfill specific design objectives \citep{Alamdari2024}. In EvoDiff, a forward process incrementally modifies a protein sequence, while a reverse process learns the rules to undo the corruption. Although the method is generally applicable for designing proteins with folded domains, it has also been used to design proteins with certain disordered properties. Specifically, when EvoDiff is trained on a set of human IDRs and MSAs of their orthologues \citep{Lu2022b}, it may be used to generate novel and diverse sequences that preserve the predicted disorder scores of the original IDRs \citep{Alamdari2024}.

A different sequence-based approach was used to design novel IDRs with specific cellular localization, such as to mitochondria or biomolecular condensates \citep{Strome2023}. Starting from a random sequence, single amino-acid substitutions were introduced to progressively match 94 sequence features that characterize naturally occurring target IDRs. Despite low sequence similarity to the targets, the designed sequences largely recapitulated the expected cellular localization in experimental tests. This indicates that the bulk biophysical properties encoded in carefully chosen sequence features can be sufficient to guide selective partitioning.

A challenging design objective is to create sequences that remain disordered in isolation but adopt specific structures upon binding to a partner protein. GADIS leverages a genetic algorithm and implicit solvent all-atom MD simulations to optimize sequences for targeted helicity in the bound state while maintaining key interactions at protein-protein interfaces and preserving the native amino acid composition \citep{Harmon2016}. We envisage that such approaches, combining sequence- and physics-based ML methods, will be instrumental in deepening our understanding of folding-upon-binding.

ML methods have also begun to tackle the challenge of designing binders for IDRs, which could help unlock the potential of these traditionally `undruggable' molecules as targets for medical interventions \citep{Xie2023}. In one approach, a contrastive learning framework based on a pLM was developed to design peptide binders against conformationally diverse targets, including IDRs. The therapeutic strategy involves fusing the peptide binder with an E3 ubiquitin ligase domain, thereby priming the target protein for degradation \citep{Bhat2025}. Peptides designed through this sequence-based pipeline successfully targeted and degraded the highly disordered fusion oncoprotein SS18-SSX1. Another framework combined deep-learning models with biophysical principles to design folded proteins that bind IDRs with high affinity and specificity \citep{Wu2024}. This method involves the assembly of recognition pockets specific for single amino acids or dipeptides, thus enabling IDRs to bind in extended conformations through multiple polar interactions. In contrast to previous works targeting compact structures, such as helices or strands, where binding is primarily driven by hydrophobic contacts, this approach focuses on more flexible binding interactions thereby broadening the scope of potential therapeutic and biotechnological applications.

\section*{Future perspectives}

ML is becoming a part of the standard computational toolbox in protein science, and is now routinely used to predict structure from sequence, design new sequences for specific structures, or to train predictive models from experimental data. These approaches have in particular leveraged large systematic databases for protein structures and the ability to extract structural and functional information from sequences and alignments. We are now beginning to see a substantial impact of these approaches in studies of sequence--ensemble--function relationships for IDRs, adopting them to the specifics of IDRs, including the need for conformational ensembles and low position-wise sequence conservation. 
We expect that ML will be useful for reconciling and interpreting heterogeneous data, for example from simulated ensembles, biophysical experiments, biochemical assays, and clinical sources, as well as for moving from binary classification tasks for IDRs towards prediction of measurable quantities. We therefore continue to need high-quality data from experiments and simulations, possibly guided by ML in data collection. In that scenario, NNs might suggest follow-up experiments/simulations that best resolve contradictions and uncertainties in the current data.

\begin{table*}[htbp]
\caption{Glossary of technical terms.}
\label{tab:1}
\begin{tabularx}{\textwidth}{lX}
    \toprule
    \textbf{Biophysics} & \\
    \midrule
     Coarse-grained model & Molecular model with reduced representation: Each coarse-grained particle represents several atoms. \\
     Force field & Set of functions and parameters describing the particles and their interactions in a simulation. \\
     Conformational ensembles & Set of three-dimensional structures representing the distribution of configurations of a protein. \\
     Folding-upon-binding & Formation of secondary structure (helix, sheet) in IDRs upon binding to a partner. \\
     Phase separation & Phase transition, whereby one (homotypic) or multiple (heterotypic) components de-mix into dense and dilute phases. \\
     \midrule
     \textbf{Machine learning} & \\
     \midrule
     Supervised learning & ML model trained against labeled data (e.g. image classification). \\
     Unsupervised learning & ML model trained to recognize patterns and structure in input data without labels (e.g. dimensionality reduction, clustering). \\
     Self-supervised learning & ML model trained against data labeled by the model itself (e.g. contrastive learning, natural language processing). \\
     Active learning & ML algorithm to iteratively generate new input data based on current data and a criterion (acquisition function). \\
     Bayesian parameter learning & Parameter optimization against data while incorporating prior knowledge via Bayes' theorem. `Bayesian optimization' can specifically refer to an algorithm for searching optimal parameters of an expensive-to-evaluate function using a surrogate model and an acquisition function, which balances exploration of uncertain regions and exploitation of likely optimal regions. \\
     Genetic algorithm & Exploration and optimization algorithm based on natural selection. The best (fittest) candidates of a population are iteratively modulated (mutation or crossover) and evaluated until convergence.  \\
     Contrastive learning & ML method to encode data into a latent space such that similar data points (e.g. closely related sequences) are mapped to nearby representations, whereas dissimilar data points are mapped to distant representations. \\
     Deep learning & ML method using neural networks (NNs) with large numbers of parameters to learn complex representations or make predictions for large datasets. \\
     Dense NN & NN with one or more hidden layers. Nodes in each layer are connected to all nodes in the previous and subsequent layers. \\
     Graph NN & NN architecture of nodes (vectors of features) interconnected  by edges to work with graph-like data structures, e.g. three-dimensional protein structures. \\
     Sequence-conditioned NN & NN whose output (e.g. a generated 3D structure) is modulated by an input amino-acid sequence. \\
     Generative modeling & ML methods that learn the distribution underlying training data in order to generate artificial data samples. \\
     Generative adversarial networks & Set of two competing NNs trained to generate realistic data similar to the training samples and to discriminate between real and generated data, respectively. \\
     Generative recurrent NN & Generative model for sequential or time-dependent output. A recurrent structure maintains information from previous output. \\
     Protein language model (pLM) & Unsupervised NN that learns the features underlying sequences. Generally uses a transformer model architecture.  \\
     Transformer model & NN used in the processing of input strings (natural language, protein sequences, etc.) which learns relationships between different parts of the input sequence via an `attention' mechanism. \\
     Diffusion model & Generative ML model mapping real data to a simple distribution via iterative noising steps. New samples are generated by drawing from the simple distribution and applying the learned `de-noising' operations. \\
     \bottomrule
\end{tabularx}
\end{table*}

\section*{Acknowledgments}

This work is a contribution from the PRISM (Protein Interactions and Stability in Medicine and Genomics) centre funded by the Novo Nordisk Foundation (to K.L.-L.; NNF18OC0033950). S.v.B. acknowledges support by the European Molecular Biology Organisation through Postdoctoral Fellowship grant ALTF 810-2022. 

\section*{Conflicts of Interest}
K.L.-L. holds stock options in and is a consultant for Peptone. The remaining authors declare no competing interests.

\FloatBarrier

\bibliographystyle{elsarticle-num}

{\footnotesize
\bibliography{references}}

\end{document}